\documentclass[12pt,prd,showkeys,amsmath,amssymb,nofootinbib]{revtex4-1}

\usepackage{amssymb}
\usepackage{amsmath}
\usepackage{amsfonts}
\usepackage{bm}
\usepackage[colorlinks=true,urlcolor=blue,linkcolor=blue,citecolor=blue]{hyperref}

\begin{document}

\title{Classical electrodynamics and spin-torsion coupling effects}

\author{Mariya Iv. Trukhanova$^{1,2}$}
\email{trukhanova@physics.msu.ru}
\author{Yuri N. Obukhov$^{1}$}
\email{obukhov@ibrae.ac.ru}
\affiliation{$^{1}$~Russian Academy of Sciences, Nuclear Safety Institute (IBRAE),\\
B. Tulskaya 52, 115191 Moscow, Russia\\
$^{2}$~Faculty of Physics, Lomonosov Moscow State University,\\
Leninskie Gory-1, 119991 Moscow, Russia}

\begin{abstract}
We study the Poincar\'e gauge theory of gravity with the most general Lagrangian quadratic in curvature and torsion, focusing on the possible interaction of the axial torsion with the electromagnetic field. From the analysis of the closed system of field equations, we identify the magnetic helicity density and the spin density of the electromagnetic field as the components of the source for the torsion field. We consider the propagation of a linearly polarized plane electromagnetic wave and demonstrate the ``dichroism'' or the polarization rotation effect under the action of external electric and magnetic fields orthogonal to the direction of wave propagation.
\end{abstract}


\maketitle

\section{Introduction}

Chiral transport phenomena are in the focus of the current research in the context of Weyl semimetals \cite{HOSUR2013857,PhysRevB.88.125105,PhysRevB.88.125110}, the chiral magnetic effect \cite{Volovik} in superfluid helium $^{3}$He-A, and the physics of heavy ion collisions \cite{KHARZEEV20161}. Of special interest are the chiral separation effect \cite{PhysRevD.72.045011} and the anomalous quantum Hall effect, which are realized in these systems without an external magnetic field \cite{PhysRevB.86.115133}. The chiral magnetic effect is manifested as the electric charge separation along the external magnetic field that is induced by the chirality imbalance \cite{PhysRevD.22.3080,KHARZEEV2014133}. On the other hand, the chiral vortical effect \cite{Abramchuk} shows up as the emergence of the fermionic axial current in the presence of vorticity induced by chirality imbalance, which was predicted for the first time by A. Vilenkin \cite{PhysRevD.22.3080}. An explicit calculation \cite{PhysRevD.98.096011} of the axial current for finite rotation and temperature in a curved spacetime was based on the assumption that the Chern-Simons current can approximate the physical axial current. The chiral magnetic current is topologically protected and hence non-dissipative even in the presence of strong interactions, for example, in the strongly coupled quark-gluon plasma, where the strong magnetic field is generated by the colliding ions \cite{KHARZEEV20161}. The mean field dynamo model was proposed \cite{PhysRevD.101.083009} to describe the generation of the strong magnetic fields in a neutron star in the context of the chiral magnetic effect.

Recently, it has been proposed that a new type of chiral current is induced by the spacetime torsion in the presence of chirality imbalance. This effect was named the ``chiral torsional effect'' \cite{Khaidukov,PhysRevD.102.016001}, and the torsion-induced chiral currents were derived \cite{PhysRevD.102.016001} in the most general gravity theory with finite temperature, density, and curvature. The interest in the theory of gravitation with spin and torsion based on the Riemann-Cartan geometry had considerably grown in the second half of the 20th century, after the consistent gauge-theoretic formalism was developed \cite{2,10.1063/1.1703702,Trautman:2006fp,PhysRevD.10.1066}. It is now well established that the spacetime torsion can only be detected with the help of the spin \cite{PhysRevD.21.2081,Hehl,Obukhov2015} and the early theoretical analysis \cite{RevModPhys.48.393,Hehl} revealed possible experimental manifestations of the torsion field. The wide class of the Poincar\'e gauge gravity models with the most general Lagrangian quadratic in the curvature and torsion was studied \cite{Poincare2} with an emphasis on the analysis of the consistency of the gauge theory of gravity with experimental observations at the macroscopic level. Accordingly, possible post-Riemannian deviations of the spacetime geometry could be essentially tested at the microscopic level, based on the study of the dynamics of fundamental particles, atoms and molecules. So far, one could not create a source of spin density that could generate torsion to be detected in a laboratory. However, one can establish the constraints on the spin-torsion coupling, in particular from the experimental search for the Lorentz and $CPT$ symmetries violation.

Although the torsion of spacetime has not yet been detected, it can be realized in condensed matter by a line defect of a crystal lattice \cite{PhysRevD.100.054509}. The geometrical formalism that associates a density of dislocations with the torsion tensor in graphene has been developed \cite{DEJUAN2010625}, and a torsion-modified anomalous Hall effect was predicted in Weyl semimetals \cite{PhysRevB.99.155152}.

Topological current can be a source of the axial torsion field \cite{TAO}. In this paper, we investigate an extended electrodynamics constructed in the framework of the Poincar\'e gauge gravity model on the background of the flat spacetime metric with nontrivial torsion, and study the effects of the chiral topological current on the torsion wave generation.

Our basic conventions and notations are as follows. The world indices are labeled by Latin letters $i,j,k,\ldots = 0,1,2,3$, while Greek letters are used for tetrad indices, $\alpha,\beta,\ldots = 0,1,2,3$ with respect to anholonomic frames. In order to distinguish separate tetrad indices we put hats over them. Finally, spatial indices are denoted by Latin letters from the beginning of the alphabet, $a,b,c,\ldots = 1,2,3$. The metric of the Minkowski spacetime reads $g_{ij} = {\rm diag}(c^2, -1, -1, -1)$, and the totally antisymmetric Levi-Civita tensor $\eta_{ijkl}$ has the only nontrivial component $\eta_{0123} = c$, so that $\eta_{0abc} = c\varepsilon_{abc}$ with the three-dimensional Levi-Civita tensor $\varepsilon_{abc}$. The spatial components of the tensor objects are raised and lowered with the help of the Euclidean 3-dimensional metric $\delta_{ab}$.

\section{Axial torsion dynamics}

The Poincar\'e gauge gravity \cite{HEHL19951,doi:10.1142/S021988780600103X,doi:10.1142/S0219887818400054} is a viable extension of Einstein's general relativity theory (GR), in which the spin, along with energy and momentum, is an independent source of the gravitational fields, and the spacetime structure is described by the Riemann-Cartan geometry with the curvature 
\begin{align}\label{curv}
R_{kli}{}^j &= \partial_k\Gamma_{li}{}^j - \partial_l\Gamma_{ki}{}^j
+ \Gamma_{kn}{}^j \Gamma_{li}{}^n - \Gamma_{ln}{}^j\Gamma_{ki}{}^n,
\end{align}
and the torsion 
\begin{align}\label{tors}
  T_{kl}{}^i &= \Gamma_{kl}{}^i - \Gamma_{lk}{}^i.
\end{align}

A general metric-affine spacetime manifold is endowed with two main geometrical structures: the metric $g_{ij}$ and linear connection $\Gamma_{ki}{}^j$. The metric determines the distances and lengths of vectors and the angles between vectors. On the other hand, the connection defines the parallel transport and covariant derivatives. From the geometrical point of view, the torsion measures the non-closure of the parallelogram obtained by means of the parallel transport of two vectors along each others direction. The Riemann-Cartan geometry arises as a special case of the metric-affine spacetime when the nonmetricity vanishes, $\nabla_k g_{ij} = 0$. 

Technically, the Riemann-Cartan connection can be decomposed
\begin{equation}
\Gamma_{kj}{}^i = \widetilde{\Gamma}_{kj}{}^i - K_{kj}{}^i,\label{dist}
\end{equation}
into a Riemannian term (Christoffel symbols), which depends on the metric, 
\begin{equation}\label{Chr}
\widetilde{\Gamma}_{kj}{}^i = {\frac 12}g^{il}(\partial_jg_{kl} + \partial_kg_{lj} - \partial_lg_{kj}),
\end{equation}
and the post-Riemannian contortion tensor that is determined by the torsion 
\begin{equation}\label{NTQ}
K_{kj}{}^i = -\,{\frac 12}(T_{kj}{}^i + T^i{}_{kj} + T^i{}_{jk})\,.
\end{equation}
Furthermore, the torsion tensor (\ref{tors}) can be decomposed \cite{SHAPIRO2002113} into the three irreducible components
\begin{equation}
T_{kl}{}^{i} = {}^{(1)}T_{kl}{}^{i} + {}^{(2)}T_{kl}{}^{i} + {}^{(3)}T_{kl}{}^{i},
\end{equation}
where the second and the third irreducible parts 
\begin{equation}\label{tT2}
{}^{(2)}T_{kl}{}^{i} = {\frac 13}\!\left(\delta^i_{k}T_{l} - \delta^i_{l}T_{k}\right),\qquad
{}^{(3)}T_{kl}{}^{i} = -\,{\frac 13}\,\eta_{kl}{}^{ij}\overline{T}{}_j, 
\end{equation}
feature the torsion trace vector and the axial torsion pseudovector, respectively,
\begin{equation}
T_j := T_{ij}{}^i,\qquad \overline{T}{}^j = {\frac 12}T_{kli}\eta^{klij},\label{Pv}
\end{equation}
whereas the first irreducible part is a totally traceless tensor
\begin{equation}
{}^{(1)}T_{ik}{}^{i} = 0,\qquad {}^{(1)}T_{ijk}\eta^{ijkl} = 0.\label{tT1}
\end{equation}

Here we focus on the dynamical realization of the Poincar\'e gauge theory as a Yang-Mills type model \cite{Poincare2} with the most general quadratic in curvature and torsion Lagrangian. Earlier \cite{Yakushin}, the contribution of the vector and the pseudovector (\ref{Pv}) to physical effects at the microscopic level was analysed in the framework of this theory, and the strong constraints were established on the spin-torsion coupling parameters. Following the previous research \cite{Yakushin}, we continue to study the influence of the {\it axial pseudovector} torsion field on physical matter, by assuming that the metric of spacetime is flat. As a result, the connection (\ref{dist}) reduces to the contortion piece, $\Gamma_{kj}{}^i = - K_{kj}{}^i$, so that the free torsion field is described by an effective Yang-Mills-type gauge gravity Lagrangian
\begin{equation}
L_T = \hbar\left\{-\,{\frac 14}f_{ij}f^{ij} + {\frac {\mu^2}2}\,\alpha_i\alpha^i
- {\frac \lambda 2}(\partial_i\alpha^i)^2\right\}.\label{LRTa}
\end{equation}
Here $f_{ij} = \partial_i\alpha_j - \partial_j\alpha_i$ is constructed from the rescaled axial torsion field
\begin{equation}\label{alpha}
\alpha_i = {\frac {\ell_\rho}{3}}\sqrt{\frac{-\Lambda_5}{2\kappa c\hbar}}\,\overline{T}_i,
\end{equation}
where Einstein's gravitational constant $\kappa$ is responsible for long-range effects (``macroscopic gravity''), the short-range behavior (``microscopic gravity'') is regulated by $\ell_\rho$ with the dimension of length, and parameters 
\begin{equation}\label{mula}
\mu^2 = -\,{\frac{3\mu_1}{\ell_\rho^2\Lambda_5}},\qquad \lambda = {\frac {3\Lambda_4}{2\Lambda_5}},
\end{equation}
are defined in terms of the coupling constants $\Lambda_4, \Lambda_5, \mu_1$ which determine the spin-parity particle spectrum of the general quadratic Poincar\'e gauge gravity model \cite{primer}. For the spin-one sector, one finds $\lambda = 0$, thus avoiding stability and unitarity problems. One can also assume that the rest mass vanishes $\mu = 0$.

\subsection{Interaction between fermions and axial torsion}

In accordance with the minimal coupling principle \cite{PhysRevD.90.124068}, the interaction between the fermion field $\psi$ and gauge fields is introduced via the spinor covariant derivative 
\begin{equation}\label{Ds}
D_i = \partial_i - {\frac {iq}{\hbar}}A_i + \frac{i}{4}\Gamma_i{}^{\alpha\beta}\sigma_{\alpha\beta},
\end{equation}
with the Lorentz group generators $\sigma^{\alpha\beta} =\frac{i}{2}(\gamma^{\alpha}\gamma^{\beta}-\gamma^{\beta}\gamma^{\alpha})$ constructed from the Dirac matrices $\gamma^\alpha$:
\begin{equation}
\gamma^0=\frac{1}{c}\begin{pmatrix} 1 & 0 \\ 0 & -1\end{pmatrix}, \qquad
\gamma^a=\begin{pmatrix} 0 & \sigma^a \\ -\sigma^a & 0 \end{pmatrix},\qquad  
\gamma_5=\begin{pmatrix} 0 & -1 \\ -1 & 0 \end{pmatrix}.
\end{equation}
Making use of the covariant derivative (\ref{Ds}) in the Lagrangian of the Dirac field $\psi$, we find that the dynamics of the latter on the Minkowski flat metric background is derived from
\begin{equation}\label{Dirac_field}
L_D = \frac{i\hbar}{2}\left\{\overline{\psi}\gamma^i\partial_i\psi -
(\partial_i\overline{\psi})\gamma^i\psi\right\} - mc\overline{\psi}\psi +
qA_i\overline{\psi}\gamma^i\psi + {\frac 34}\hbar\chi\alpha_i\overline{\psi}\gamma^i\gamma_5\psi,
\end{equation}
where $\overline{\psi}$ is the Dirac conjugate spinor, and we used the identity $\gamma^{\mu}\sigma^{\alpha\beta} + \sigma^{\alpha\beta}\gamma^{\mu} =-\,2\eta^{\mu\nu\alpha\beta}\gamma_{\nu}\gamma_{5}$, with $\eta^{0abc} = -\,{\frac{1}{c}}\epsilon^{abc}$. The interaction between the Dirac fermion and the axial field (\ref{alpha}) is determined by the dimensionless coupling constant
\begin{equation}\label{chi}
\chi = {\frac{1}{\ell_\rho}}\sqrt{\frac {2\kappa c\hbar}{-\Lambda_5}}
= {\frac{\ell_{\rm Pl}}{\ell_\rho}}\sqrt{\frac {16\pi}{-\Lambda_5}}.
\end{equation}
The last term in the Lagrangian (\ref{Dirac_field}) features the coupling of the axial torsion to the spin and helicity density of the Dirac fermion. The spin-torsion coupling constant $\chi$ is {\it very small}, provided we assume that the characteristic length of the Poincar\'e gauge gravity is much larger than the Planck scale, $\ell_\rho\gg\ell_{\rm Pl}$.

\subsection{Interaction between the electromagnetic field and axial torsion}

The interaction of the axial torsion and the electromagnetic field is derived from the standard Maxwell-Lorentz Lagrangian when the ordinary derivatives are replaced by covariant ones $D_iA^j=\partial_iA^j+\Gamma_{ik}^{\ \ j}A^k=\partial_iA^j-\frac{1}{6}\eta_{ik}^{\ \ jl}A^k\overline{T}_{l}$. An apparent breaking of the $U(1)$ gauge invariance can be fixed by a modified Stueckelberg's method \cite{Stueckelberg}. As a result, the interaction between the electromagnetic field and the axial torsion can be derived from the Lagrangian
\begin{equation}
L_I = -\,{\frac{1}{4}}\sqrt{\frac{\varepsilon_0}{\mu_0}}F_{ij}F^{ij}
+ \chi\sqrt{\frac{\varepsilon_0}{\mu_0}}\alpha_i\eta^{ijkl}A_j\partial_kA_l,
\end{equation}
where $\varepsilon_0$ and $\mu_0$ are the electric and magnetic constants of the vacuum. Together with the dynamical Lagrangian for the axial vector field (\ref{LRTa}) and the fermion sector terms (\ref{Dirac_field}), the total Lagrangian for the torsion field interacting with spin of the matter sources has the form
\begin{eqnarray}
L &=& -\,\frac{1}{4}\sqrt{\frac {\varepsilon_0}{\mu_0}}\,F_{ij}F^{ij} + \frac{i\hbar}{2}\left\{
\overline{\psi}\gamma^i\partial_i\psi - (\partial_i\overline{\psi})\gamma^i\psi\right\}
- mc\overline{\psi}\psi + qA_i\overline{\psi}\gamma^i\psi \nonumber\\
&& +\,\hbar\Bigl\{-\,{\frac 14}f_{ij}f^{ij} + {\frac {\mu^2}2}\,\alpha^2 - {\frac \lambda 2}
(\partial\alpha)^2\Bigr\} \nonumber\\
&& + \,{\frac 34}\hbar\chi\alpha_i\overline{\psi}\gamma^i\gamma_5\psi +
\chi\sqrt{\frac {\varepsilon_0}{\mu_0}}\,\alpha_i\eta^{ijkl}A_j\partial_kA_l.\label{Lagrangian1}
\end{eqnarray}
The last line describes the coupling of axial torsion field to the axial current density of the Dirac fermion and the photon fields:
\begin{equation}\label{current}
j^i_f = \overline{\psi}\gamma^i\gamma_5\psi,\qquad j^i_b = \eta^{ijkl}A_j\partial_kA_l,
\end{equation}
where we have explicitly
\begin{equation}\label{AA}
A_i = \{-\,\phi, \bm{A}\}, \qquad \alpha_i = \{-\,\varphi, \bm{\alpha}\}.
\end{equation}
Earlier \cite{Yakushin}, the quantization of the model (\ref{Lagrangian1}) was analysed, and the static potential between fermions due to the exchange of the axial torsion was computed.

\subsection{Field equations}

Neglecting the fermion sector, let us derive the field equations for the Lagrangian (\ref{Lagrangian1}). Technically, we need to make variations with respect to the axial torsion field $\alpha_i$ and with respect to the electromagnetic field potential $A_i$. The corresponding Euler-Lagrange equation $\delta L/\delta\alpha_i = 0$ for the torsion reads:
\begin{equation}\label{eqT}
-\,\partial_jf^{ij} + \mu^2\alpha^i + \lambda\partial^i(\partial\alpha)
+ \frac{\chi}{\hbar}\sqrt{\frac {\varepsilon_0}{\mu_0}}\,\eta^{ijkl}A_j\partial_kA_l = 0.
\end{equation}
Technically, it will be convenient to proceed from the four-dimensional language to the 3-vector description by introducing the three-component fields
\begin{equation}\label{EBt}
{\mathcal E}_a = f_{a0},\qquad {\mathcal B}^a = {\frac 12}\epsilon^{abc}f_{bc},\qquad a,b = 1,2,3, 
\end{equation}
as the ``electric'' and ``magnetic'' strengths for the scalar and vector potentials (\ref{AA}), $\varphi$ and $\bm{\alpha}$, respectively: $\bm{\mathcal E} = -\,\bm{\nabla}\varphi - \partial_t\bm{\alpha}$, and $\bm{\mathcal B} = \bm{\nabla}\times\bm{\alpha}$. Recall the usual definitions for the electric and magnetic fields:
\begin{equation}\label{EBm}
 E_a = F_{a0},\qquad B^a = {\frac 12}\epsilon^{abc}F_{bc},\qquad a,b = 1,2,3.
\end{equation}
We then recast (\ref{eqT}) into the 3-dimensional form:
\begin{eqnarray}
\bm{\nabla}\cdot\bm{\mathcal E} + \mu^2\varphi - \lambda\partial_t(\partial\alpha) &=& 
-\,{\frac{\chi}{\mu_0\hbar}}\,\bm{A}\cdot\bm{B},\label{eqT1}\\
\bm{\nabla}\times\bm{\mathcal B} -{\frac 1{c^2}}\partial_t\bm{\mathcal E} + \mu^2\bm{\alpha}
+ \lambda\bm{\nabla}(\partial\alpha) &=& -\,\frac{\chi}{\hbar}\varepsilon_0\left(\phi\bm{B} 
+ \bm{E}\times\bm{A}\right).\label{eqT2}
\end{eqnarray}
The first equation (\ref{eqT1}) is an extended analogue of Gauss's law for the ``electric'' field $\bm{\mathcal E}$. The source on the right side of this equation is the helicity density, which is determined by the scalar product of the electromagnetic vector potential and the magnetic field $\sim \bm{A}\cdot\bm{B}$. The magnetic helicity is an important quantity in the study of topological configurations of electric and magnetic lines, that can be understood as an average measure of how much the field lines are knotted and linked. In a similar way, knots can be realized in fluid dynamics \cite{MAGGIONI201329}, optics \cite{Zhong:21} and liquid crystals \cite{Smalyukh_2020}. Knotted structures in hydrodynamical fields, for example, are realized in magnetic field lines of a plasma \cite{PhysRevLett.115.095001} or vortex lines of classical and quantum fluids. Such a field configuration is topologically nontrivial and can demonstrate Hopf links and Hopf fibration. 

An exact analogue of Gauss's law arises for the special class of stable and unitary models with $\lambda=0$ and the massless axial field $\mu=0$:
\begin{equation}
\oint \bm{\mathcal E}\cdot d\bm{\sigma} = - \,\frac{\chi}{\mu_0\hbar}\int d^3x \bm{A}\cdot\bm{B}.
\end{equation}
Interestingly, the conservation equation for the magnetic helicity density can be derived from the Maxwell equations when $\bm{E}\cdot\bm{B}=0$:
\begin{equation}
\partial_t(\bm{A}\cdot\bm{B})+\bm{\nabla}\cdot\left(\phi\bm{B}+\bm{E}\times\bm{A}\right) = 0.
\end{equation}

The second field equation (\ref{eqT2}) is an analogue of Ampere's law for the ``magnetic'' field $\bm{\mathcal B}$, where the spin density of the electromagnetic field $\sim \phi\bm{B}+\bm{E}\times\bm{A}$ appears as the corresponding source on the right-hand side.

In order to obtain the closed system, we derive the modified Maxwell equations from the variation of the Lagrangian (\ref{Lagrangian1}) with respect to the 4-potential of the electromagnetic field $A_i$: 
\begin{equation}\label{eqM}
-\,\partial_jF^{ij} + {\frac{\chi}{2}}\,\eta^{ijkl}A_jf_{kl} - \chi\,\eta^{ijkl}\alpha_jF_{kl} = 0.
\end{equation}

An immediate observation is in order. Since $\partial_i\partial_jF^{ij} = 0$ identically (symmetric lower indices contracted with the antisymmetric upper indices), by taking the divergence $\partial_i$ of the field equations (\ref{eqT}) [in the special class of models with $\mu^2 = 0$ and $\lambda = 0$] and (\ref{eqM}), we derive
\begin{equation}\label{FF0}
\eta^{ijkl}F_{ij}F_{kl} = 0,\qquad \eta^{ijkl}F_{ij}f_{kl} = 0,
\end{equation}
respectively. Making use of (\ref{EBt}) and (\ref{EBm}), we thus find that only crossed-field configurations are actually allowed:
\begin{equation}\label{EB0}
\bm{E}\cdot\bm{B} = 0,\qquad \bm{E}\cdot\bm{\mathcal B} + \bm{\mathcal E}\cdot\bm{B} = 0.
\end{equation}
In particular, this includes wave configurations. 

By making use of (\ref{AA}), (\ref{EBt}) and (\ref{EBm}), we rewrite the inhomogeneous Maxwell equations (\ref{eqM}) in the 3-dimensional form:
\begin{align}
\bm{\nabla}\cdot\bm{E} &= 2\chi c\,\bm{\alpha}\cdot\bm{B}
- \chi c\,\bm{A}\cdot\bm{\mathcal B},\label{eqM1}\\
\bm{\nabla}\times\bm{B} - {\frac 1{c^2}}\partial_t\bm{E} &= 
{\frac{2\chi}{c}}\left(\varphi\bm{B} + \bm{E}\times\bm{\alpha}\right) -
{\frac{\chi}{c}}\left(\phi\bm{\mathcal B} + \bm{\mathcal E}\times\bm{A}\right).\label{eqM2}
\end{align}
As usual, we have to add the homogeneous Maxwell system,
\begin{equation}
\bm{\nabla}\cdot\bm{B} = 0,\qquad
\bm{\nabla}\times\bm{E} + \partial_t\bm{B} = 0.\label{eqMh2}
\end{equation}

\section{Dichroism in an external electromagnetic field}

Let us consider a region of space filled by the homogeneous dielectric matter with the relative permittivity $\varepsilon$. In the presence of the homogeneous static magnetic and electric fields in this region, the photon-torsion dynamics is described by the modified Maxwell's equations
\begin{eqnarray}\label{Eq1}
\varepsilon\bm{\nabla}\cdot\bm{E} &=& 2\chi c \bm{\alpha}\bm{B}_0,\\ 
\bm{\nabla}\times\bm{B} - \frac{\varepsilon}{c^2}\partial_t\bm{E}
&=& \frac{2\chi}{c}\left(\varphi\bm{B}_0+\bm{E}_0\times\bm{\alpha}\right),
\end{eqnarray}
and the equations of motion for the axial torsion field potentials
\begin{eqnarray}
\left(\frac{1}{c^2}\partial^2_t-\nabla^2+\mu^2\right)\varphi
&=& -\,{\frac{\chi}{\mu_0\hbar}}\,\bm{A}\cdot\bm{B}_0,\label{Eq2}\\
\left(\frac{1}{c^2}\partial^2_t-\nabla^2+\mu^2\right)\bm{\alpha}
&=& -\,{\frac{\chi\varepsilon_0}{\hbar}}\,\bm{E}_0\times\bm{A},\label{Eq3}
\end{eqnarray}
where we use the gauge $\bm{\nabla}\cdot\bm{A} = 0$ and $\phi=0$, and consider the class of models with $\lambda = 0$ \cite{Yakushin2}.

\subsection{External fields orthogonal to the wave propagation}

Let us analyze the wave propagation along $\bm{e}_z$ in the presence of the external electric $\bm{E}_0=E_0\bm{e}_x,$ and magnetic $\bm{B}_0=B_{0}\bm{e}_{x}$ fields which are thus parallel to each other and orthogonal to the wave propagation $\bm{k}=k\bm{e}_z$. We look for solutions of Eqs. (\ref{Eq1})-(\ref{Eq3}) as a linear superposition of plane waves
\begin{equation}
\bm{A}(z,t) = \widetilde{\bm{A}}\,e^{-i\omega t + ikz}, \qquad
\bm{\alpha}(z,t) = \widetilde{\bm{\alpha}}\,e^{-i\omega t + ikz},
\end{equation}
and substituting this ansatz into (\ref{Eq1})-(\ref{Eq3}), we then derive $A_z = 0$, $\alpha_x = \alpha_y = 0$, and
\begin{eqnarray}
\left(k^2 - \frac{\varepsilon\omega^2}{c^2}\right)\begin{pmatrix} A_x \\ A_y\end{pmatrix}
&=& \frac{2\chi}{c}\begin{pmatrix} B_0\varphi \\ - E_0\alpha_z\end{pmatrix},\\
\left(k^2 - \frac{\omega^2}{c^2}+\mu^2\right)\begin{pmatrix}\alpha_z \\ \varphi\end{pmatrix}
&=& -\,{\frac{\chi \varepsilon_0}{\hbar}}\begin{pmatrix} E_0A_y \\ c^2B_0A_x\end{pmatrix}.
\end{eqnarray}
As we can see from this set of equations, $A_{x}, A_y$ modes oscillate into the axial torsion field components $\varphi$ and $\alpha_z$, and vice versa. The above linearized system of equations leads to the dispersion relations
\begin{equation}
k^2_{{B,E}_\pm} = {\frac{1}{2}}\left((\varepsilon+1)\frac{\omega^2}{c^2} - \mu^2
\pm \sqrt{\left((\varepsilon - 1){\frac{\omega^2}{c^2}} + \mu^2\right)^2 + 4\xi_{B,E}}\right),
\end{equation}
where we denoted $\xi_B = -\,{\frac{2\chi^2 B_0^2}{\mu_0\hbar c}}$, and $\xi_E = {\frac{2\chi^2 \varepsilon_0E_0^2}{\hbar c}}$. As a result, under the assumption $\frac{\chi^2 \varepsilon_0E_0^2}{\hbar c} \ll \frac{\omega^4}{c^4}$ and $\frac{\chi^2 B_0^2}{\mu_0\hbar c} \ll \frac{\omega^4}{c^4}$, in the first order approximation in the small coupling constant $\chi^2$, we obtain the wave numbers 
\begin{eqnarray}
k^2_{B,E_+} &=& \frac{\varepsilon\omega^2}{c^2}
+ \frac{\xi_{B,E}}{(\varepsilon - 1)\frac{\omega^2}{c^2}+\mu^2},\\
k^2_{B,E_-} &=& \frac{\omega^2}{c^2} - \mu^2
- \frac{\xi_{B,E}}{(\varepsilon - 1)\frac{\omega^2}{c^2}+\mu^2}.
\end{eqnarray}
The general solution can thus be written as 
\begin{equation}
\begin{pmatrix} A_x \\ A_y \\ \alpha_z \end{pmatrix} = e^{-i\omega t + ik_{+}z}
\begin{pmatrix} A_{x} \\ A_y \\ \alpha_z \end{pmatrix}_{+} + e^{-i\omega t + ik_{-}z}
\begin{pmatrix} A_{x} \\ A_y \\ \alpha_z \end{pmatrix}_{-}\,.
\end{equation}
For a wave traveling in the $z$ direction, let us assume that at $z=0$ we have a purely electromagnetic wave linearly polarized at an angle $\beta_0$ with respect to $\bm{e}_x$:
\begin{equation}
A_x(0,t) = {\mathcal A}\,\cos\beta_0\,e^{-i\omega t}, \quad
A_y(0,t) = {\mathcal A}\,\sin\beta_0\,e^{-i\omega t},\quad \varphi(0,t)=\bm{\alpha}(0,t) = 0.
\end{equation}
As a result, we derive a general solution for the vector potential of the electromagnetic wave field:
\begin{eqnarray}
A_x(z,t) &=& Re\left[\frac{{\mathcal A}\,\cos\beta_0}{\gamma^2 + 2\xi_B}\,e^{-i\omega t}\left(
{\xi_B}\,e^{ik_{B_-}z} + (\gamma^2 + \xi_B) e^{ik_{B_+}z}\right)\right],\label{Ax}\\
A_y(z,t) &=& Re\left[\frac{{\mathcal A}\,\sin\beta_0}{\gamma^2 + 2\xi_E}\,e^{-i\omega t}\left(
{\xi_E}\,e^{ik_{E_-}z} + (\gamma^2 + \xi_E) e^{ik_{E_+}z}\right)\right],\label{Ay}
\end{eqnarray}
where we denoted $\gamma(\omega)=(\varepsilon - 1)\frac{\omega^2}{c^2}+\mu^2.$ We expect that the magnitude of $x$ and $y$ components of the light wave should decrease in the process of conversion of the photons into the torsion waves. One can recast solutions into the form
\begin{equation}
A_{x,y}(z,t) \sim e^{- i\omega t + ik_0z}\left(1 - \frac{2\xi_{B,E}}{\gamma^2}\sin^2({qz}/{2})    
+\frac{i\xi_{B,E}}{\gamma}\left(\frac{z}{2k_0} - \frac{\sin(qz)}{\gamma}\right)\right),
\end{equation}
where $q(\omega)=\sqrt{\varepsilon}\frac{\omega}{c} - \sqrt{\frac{\omega^2}{c^2}-\mu^2}$ and $k_0 = \sqrt{\varepsilon}\frac{\omega}{c}$. The polarization of the light wave rotates, and at $z=l$ we find for the angle $\beta$ between the polarization vector and $\bm{e}_x$:
\begin{equation}\label{beta}
\tan\beta = \left(1 - \frac{8\chi^2\rho_0}{\gamma^2\hbar c}
\sin^2({ql}/{2})\right)\tan\beta_0,
\end{equation}
where $\rho_0 = \frac{\varepsilon_0E_0^2}{2} + \frac{B_0^2}{2\mu_0}$ is the energy density, and $\beta = \beta(\omega)$ depends on the frequency $\omega$ of an electromagnetic wave.

In addition, for the spatial component of the axial torsion potential we find 
\begin{equation}
\alpha_z = -\,Re\left[{\frac{{\mathcal A}\,\sin\beta_0\,e^{-i\omega t}\,\chi\,\varepsilon_0 E_0}
{({k^2_{+} - k^2_{-}})\,\hbar}}\left(e^{ik_{+}z} - e^{ik_{-}z}\right)\right].
\end{equation}
The solution for $\bm{\alpha}=\alpha_z\bm{e}_z$ is non-trivial and it corresponds to a very peculiar case of longitudinal waves, for which $\bm{\mathcal B}=\bm{\nabla}\times\bm{\alpha}=0$, and only the ``electric'' component $\bm{\mathcal E}$ of the torsion field is generated and propagates in $z$ direction.

It is well known that a linearly polarized light plane wave passing through an external constant magnetic field directed orthogonally to its propagation acquires ellipticity, thus featuring the dichroism effect, or the plane of its polarization rotation \cite{RevModPhys.93.015004}. It occurs as a result of the photon-axion conversion in a magnetic field, which is described by the modified Maxwell's equations. Here we demonstrate the existence of a similar dichroism effect (\ref{beta}) due to a conversion of the photons into the axial torsion waves, for the light propagating through the region of space with external magnetic and electric fields.

\section{Conclusions}

New experimental approaches are needed to search for the hypothetical particles and fields. Here we have proposed a possible experimental setup, within the framework of the Poincar\'e gauge theory of gravity, suitable for the study of the interaction of the axial torsion field with the electromagnetic field in a flat spacetime metric. The closed system of field equations (\ref{eqT1}), (\ref{eqT2}), (\ref{eqM1}), (\ref{eqM2}) was derived, revealing the density of magnetic helicity and the density of the spin of the electromagnetic field as the sources of the dynamic torsion field. We have focused on the study of physical effects arising due to coupling of the axial torsion with the axial photon current in the Lagrangian density (\ref{Lagrangian1}). We have considered the possibility of converting electromagnetic waves (photons) into the torsion waves when a linearly polarized plane electromagnetic wave propagates through the region of space in external electric and magnetic fields. In the regime when these fields are uniform and directed orthogonally to the wave propagation direction, the demonstrate the effect of rotation of wave's polarization. The explicit solutions (\ref{Ax}) and (\ref{Ay}) show that the polarization rotation angle $\beta(\omega)$ depends on the frequency $\omega$ of the wave (\ref{beta}), being proportional to the square of the torsion coupling constant, $\chi^2$. 

The new effect can be used for the search of possible post-Riemannian deviations of the spacetime geometrical structure to improve the experimental estimates on the magnitude of the spacetime torsion \cite{Lamm,Alan,PhysRevD.90.124068,TAO}.

\section*{Acknowledgments}

The research of Mariya Iv. Trukhanova is supported by the Russian Science Foundation under the grant No. 22-72-00036 (\url{https://rscf.ru/en/project/22-72-00036/}).

\end{document}